# MSHT: Multi-stage Hybrid Transformer for the ROSE Image Analysis of Pancreatic Cancer

Tianyi Zhang, Yunlu Feng, Yu Zhao, Guangda Fan, Aiming Yang, Shangqin Lyu, Peng Zhang, Fan Song, Chenbin Ma, Yangyang Sun, Youdan Feng, and Guanglei Zhang, *Member, IEEE*

***Abstract*—Pancreatic cancer is one of the most malignant cancers in the world, which deteriorates rapidly with very high mortality. The rapid on-site evaluation (ROSE) technique innovates the workflow by immediately analyzing the fast stained cytopathological images with on-site pathologists, which enables faster diagnosis in this time-pressured process. However, the wider expansion of ROSE diagnosis has been hindered by the lack of experienced pathologists. To overcome this problem, we propose a hybrid high-performance deep learning model to enable the automated workflow, thus freeing the occupation of the valuable time of pathologists. By firstly introducing the Transformer block into this field with our particular multi-stage hybrid design, the spatial features generated by the convolutional neural network (CNN) significantly enhance the Transformer global modeling. Turning multi-stage spatial features as global attention guidance, this design combines the robustness from the inductive bias of CNN with the sophisticated global modeling power of Transformer. A dataset of 4240 ROSE images is collected to evaluate the method in this unexplored field. The proposed multi-stage hybrid Transformer (MSHT) achieves 95.68% in classification accuracy, which is distinctively higher than the state-of-the-art models. Facing the need for interpretability, MSHT outperforms its counterparts with more accurate attention regions. The results demonstrate that the MSHT can distinguish cancer samples accurately at an unprecedented image scale, laying the foundation for deploying automatic decision systems and enabling the expansion of ROSE in clinical practice. The code and records are available at: https://github.com/sagizty/Multi-Stage-Hybrid-Transformer.**

## I. INTRODUCTION

THE pancreatic cancer is one of the most highly malignant tumors of the digestive system, with a very low 5-year survival rate of about 10% [1][2]. Due to the lack of symptoms and the shortage of proper screening technology, patients diagnosed with pancreatic cancer generally present in the advanced stage, which brings striking challenges for the treatment of the patients [3]. Therefore, if diagnosed earlier before the metastasis of cancer cells, it would have a better prognosis and a higher survival rate [4].

Endoscopic ultrasonography-guided fine-needle aspiration (EUS-FNA) has achieved a remarkable diagnostic accuracy for pancreatic cancer and is now widely used [5][6]. With the utilization of endoscopic devices and sampling techniques, the current diagnostic sensitivity of pancreatic cancer is 90% to 98% and the specificity is 95% to 100% [7][8]. The relatively low diagnostic sensitivity is mainly caused by the inadequacy of the pancreatic pathology sampling. To reduce the risk of complications and the pain of the patient during puncture surgery, the times of the needle punctures are generally limited, which may lead to the missed diagnosis. Such limitations encourage researchers to make efforts in the improvement of the sampling procedure [9].

Rapid on-site evaluation (ROSE), which refers to the real-time cytopathological evaluation during the FNA procedure, has been widely used with the expectation to decrease the diagnostic period with fewer needle punctures and increase the sample adequacy [10][11]. The availability of ROSE may reduce the inadequate sample rate of pancreatic cancer by 10%-18% [12] and thus avoid the missed diagnosis during the EUS-FNA procedure. However, the core difficulty has been pointed out to be limited pathologist staffing in [13][14] which reported that the ROSE was available in only 48% and 55% of European and Asian centers. In comparison, it is available in nearly 98% of the USA centers. To expand ROSE diagnosis worldwide, automating the workflow is highly in need.

With the development of computer-aided diagnosis (CAD) techniques and the increase of computing power, artificial intelligence has played an important role in health care. The analysis of cytopathological images with deep learning technology has been widely reported with promising diagnostic accuracy in analyzing breast cancer, cervical cancer, and gastric cancer [15-20]. Although deep learning technology shows

This work is partially supported by the National Natural Science Foundation of China (No. 61871022), the Beijing Natural Science Foundation (No. 7202102), and the 111 Project (No. B13003).

T. Zhang, Y. Feng and Y. Zhao contributed equally to this work. Corresponding Author: Guanglei Zhang (e-mail: guangleizhang@buaa.edu.cn)

T. Zhang and G. Zhang is with the Beijing Advanced Innovation Center for Biomedical Engineering, School of Biological Science and Medical Engineering, Beihang University, Beijing, 100191, China (e-mails: {zhangtianyi, guangleizhang}@buaa.edu.cn)

Y. Feng and A. Yang is with the Department of Gastroenterology, Peking Union Medical College Hospital, Beijing, 100006, China (e-mails: yunluf@icloud.com, yangaiming@medmail.com.cn)

Yu Zhao is with the Department of Pathology, Peking Union Medical College Hospital, Beijing, 100006, China (e-mails: rain986532@126.com)

S. Lvu is with the School of Electronics & Computer Science, University of Southampton, Southampton, SO17 1BJ, United Kingdom (e-mails: lvshangqing@bupt.cn)

G. Fan, P. Zhang, F. Song, C. Ma, Y. Sun, Y. Feng are with the School of Biological Science and Medical Engineering, Beihang University, Beijing, 100191, China (e-mails: {a641261717, pengzhang, fansong, machenbin, syyzbh, emilyfeng}@buaa.edu.cn)

excellent potential in the diagnosis of pancreatic cancer along with the clinical innovation ROSE technique, very few works have been reported in recent years.

A deep-learning-based CAD system in ROSE of EUS-FNA specimen was firstly introduced by Hashimoto et al. [21] in 2018, in which the authors created a stage-by-stage deep learning system that achieved the sensitivity, specificity, and accuracy all at 80% on a private dataset of 450 ROSE images. In 2020, the following work by the same group [22] achieved the accuracies of 93% and 89% in diagnosing pancreatic cancer with two deep learning models called ImageNet-CNNbn and RetinaNet on a larger dataset of 1440 ROSE images. In general, the limited accuracy of the related works may be linked to the following reasons. On the one hand, the scarcity of the ROSE images due to the time-consuming workflow and the scarcity of experienced pathologists both make dataset construction an arduous task. On the other hand, ROSE images usually contain complex backgrounds, such as the presence of noisy areas and perturbations like red cells, fibers, and vacuoles, which demand a robust feature extractor for better analysis [16].

Both of the related works by Hashimoto [21][22] applied convolutional neural networks (CNN) in the analysis of ROSE images. The early stage of CNN usually has more local attention biases, capturing local features and patterns. However, due to the limited kernel size of CNN, only the spatial related pixels can be considered in a single layer. This character is beneficial to the generalization ability of the classification task, but it may limit the global modeling ability in analyzing ROSE images. Meanwhile, although the reception field of CNN models could be made larger when designed with deeper layers, the networks still easily focus on local features due to their inductive bias instead of global features [15]. Thus, it is challenging to achieve an optimized classification performance for ROSE images.

Attention-based methods shed light on the improvement of deep learning since the introduction of Transformer [23]. To enhance the global modeling ability, the encoder-based Transformer was introduced into the field of computer vision by Dosovitskiy et al. [24]. Due to its global receptive field and long-distance modeling strength, it outperforms a series of state-of-the-art CNN models [25-33]. The utilization of Transformer blocks in the analysis of medical images was gradually studied. Frontier work GasHis-Transformer [34] model, which was based on Inception [29] and BoT [35] networks, has achieved a state-of-the-art result of 96.8% in accuracy for diagnosing gastric cancer. In the research of cytopathological image analysis with Transformer structure, cell-DETR [36], which was based on DETR [37] structure, was introduced for instance segmentation on cells. However, the Transformer-based models require a large-scale dataset to constrain and fully perform their self-attention ability (Dosovitskiy et al. [24]), which makes the combination of Transformer networks with ROSE image analysis of pancreatic cancer a challenging task.

To address the issue mentioned above and construct a classification model with remarkable accuracy for ROSE image analysis, the hybrid idea between the Transformer blocks and the CNN blocks was introduced in this study. When the CNN backbone goes deeper, different stages provide various grained features of the pancreatic cells, which encode the abstractive information through different scales. Additionally, the CNN structure provides a robust feature-extracting method by its small parameter design and the character of inductive bias, which could achieve better convergence on the limited dataset. Furthermore, due to the spatial difference among the cells and their global distribution features, which are the morphological characteristics to distinguish cancer cells from normal ones, the self-attention mechanism of the Transformer is introduced to process the extracted spatial features at a global scale. The difficulty falls on how to encode the features from CNN in the global modeling process of the Transformers, in which the focus guided decoder (FGD) structure is specially designed for converting multi-stage CNN features into multi-stage attention guidance. With such hybrid construction, CNN can provide a robust fine-grained feature extractor towards the perturbation of noisy areas existing in the ROSE images and provide supportive features across different scales. At the same time, the attention guidance enables the Transformer structure to globally model the relevant information of long-distance regions without missing prominent local features.

In general, we proposed a multi-stage hybrid Transformer (MSHT) model with the combination of CNN backbone and focus guided decoder (FGD) structure in this work. The CNN is firstly used to generate different scales of features from the ROSE images, which can represent the abstractive spatial features of the cells. Then the focus blocks of the FGD structure encode the CNN feature maps into feature sequences to carry the attention information from early stages and use them as the attention guidance. Moreover, the Transformer decoder of the FGD structure is designed to globally model the deeper patterns of the feature sequences by its global receptive field, which sophisticatedly decodes the spatial attention biases into the global modeling process. Finally, a multi-layer perceptron (MLP) is used to classify the images by taking out the class token in the Transformer outputs.

The contributions of the MSHT are as follows:

(1) As one of the first works in ROSE image analysis, the proposed MSHT firstly introduces the cutting-edge Transformer blocks from the computer vision field, which improves the classification performance by the enhanced global modeling ability of the Transformer.

(2) Taking full advantage of the unique multi-stage hybrid architecture of the Transformer and CNN, MSHT is empowered to be able to extract multi-level features across different scales and thus can achieve much higher classification accuracy by using different grained features and comprehensive distribution patterns of the pancreatic cells.

(3) To leverage the spatial attention biases from CNN in the Transformer global modeling process, the FGD focus block is designed to integrate spatial features as the guidance in global modeling. The MSHT is evaluated on the most extensive ROSE dataset to our knowledge, which presents the remarkable performance and solid interpretability of our method. Meanwhile, this innovation could shed light on cytopathology and histopathology, where cells and tissues share distinctive features by their spatial changes and global relevance.

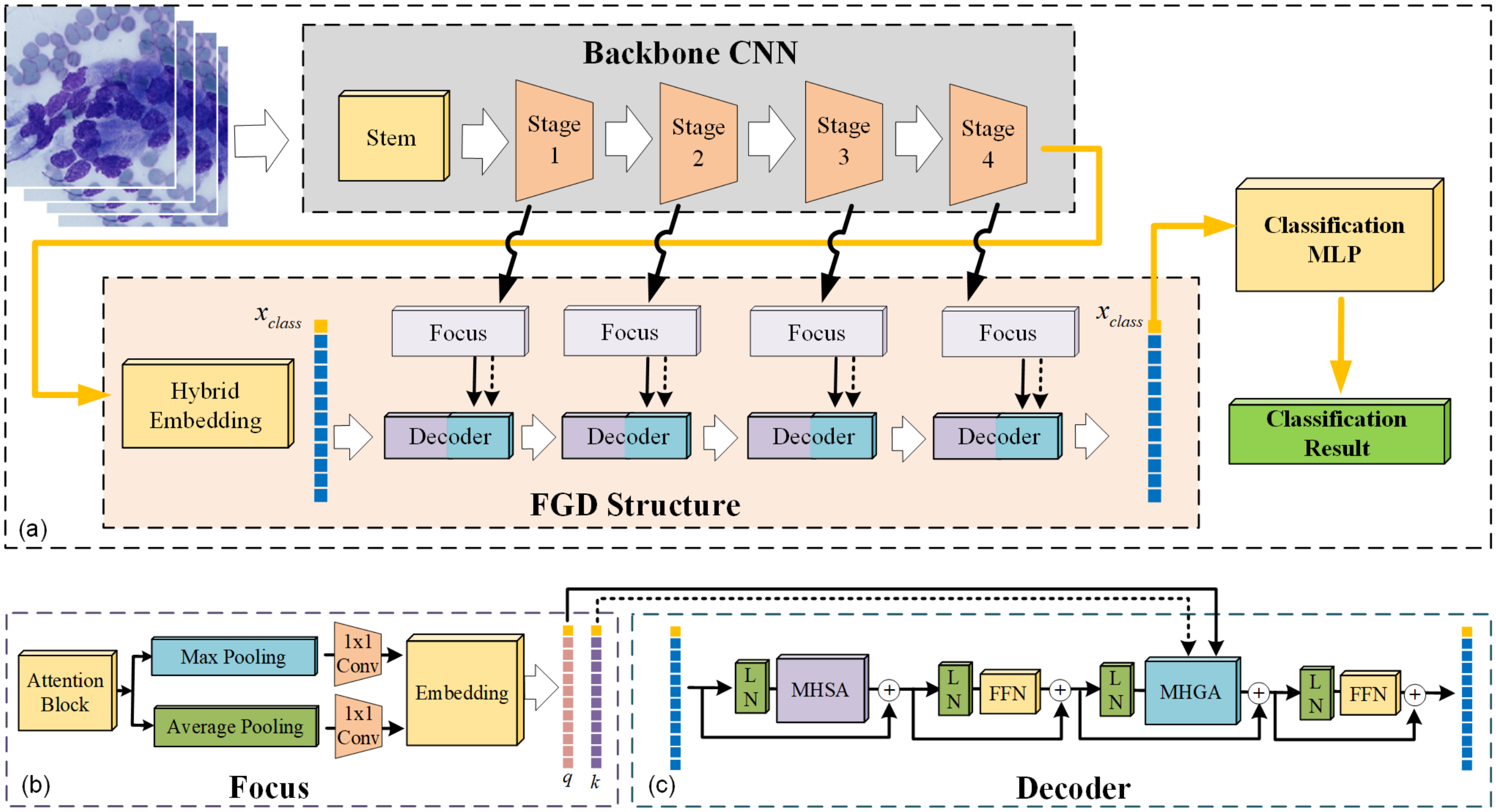

Fig. 1. The architecture of the proposed Multi-stage Hybrid Transformer (MSHT) model for the classification of ROSE images of pancreatic cancer. (a) The architecture of MSHT. (b) The focus block of the FGD structure (c) The decoder of the FGD structure. MHSA denotes multi-head self-attention, MHGA denotes multi-head guided-attention, LN denotes layer norm block, FFN denotes the feed-forward network, and MLP denotes multi-layer perceptron.

## II. Methods

The proposed Multi-stage Hybrid Transformer (MSHT) is designed for the analysis of cytopathological ROSE images of pancreatic cancer. Along with the clinical innovation strategy of ROSE, MSHT aims to diagnose pancreatic cancer faster and without time occupation of the pathologists. As shown in Fig. 1, in MSHT, the CNN backbone generates the feature maps from different stages. A FGD structure is explicitly designed for global modeling and local attention information fusion. For better global modeling, the MSHT has also equipped a novel Transformer module, which combines the attention biases of different stages of the CNN backbone through the FGD structure.

### A. Backbone CNN: ResNet50

The MSHT model uses the ResNet50 [28] as our backbone to extract features from the input ROSE images. The ResNet50 is designed by stacking a Stem block and four stages of CNN bottleneck blocks, where the blocks downsample the images into abstractive features. To effectively fuse the multi-stage features from the CNN backbone with the Transformer decoders, we modified the original backbone ResNet50 by taking out the feature maps of 4 stages and connecting them to the focus blocks of the FGD structure. Structurally, the calculation flow is reserved to maintain the mainstream feature extraction, while the last stage feature maps serve as the first inputs of the FGD structure for more complex global modeling.

### B. Focus Guided Decoder (FGD) Module

#### 1) Embedding module and hybrid embedding block

As shown in Fig. 1(a), at the beginning of the FGD structure, the hybrid embedding block transposes the feature maps of the last CNN stage into feature patches for global modeling using the Transformer blocks. And Fig. 2 illustrates the calculation process of the hybrid embedding block, which is combined with a patch projection and an embedding process.

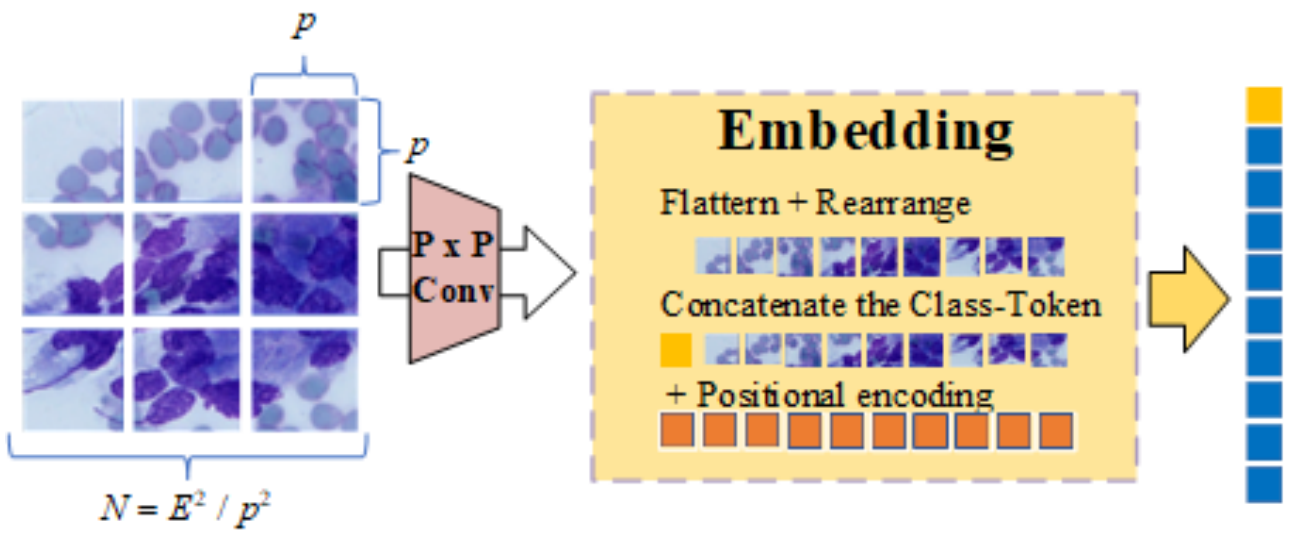

Fig. 2. The architecture of the hybrid embedding block.

During the patch projection process, a single CNN layer is used to split the input feature maps into $N = E^2 / p^2$ patches by setting its kernel size and stride, both equal to the same patch size $p$. Meanwhile, the channel setting of the CNN layer needs to project the dimension of the CNN feature maps $C$ to match the input dimension of the Transformer $D$. After the CNN projection, the feature maps are transformed into feature sequences through the embedding module to meet the prerequisites of the Transformer. The class token and positional encoding are used in all embedding processes, which is hired in both the hybrid embedding block and the focus block.

The class token $x_{class}$ is a learnable matrix that serves as the first token in Vision Transformers [24]. In MSHT, a global shared parameter patch $x_{class} \in \mathbf{R}^{1\times D}$ is deployed to carry the classification information throughout the FGD structure, as shown in Fig. 1(a). Initially, it is an empty token represented by a zero tensor $x_{class} \in O^{1\times D}$. In the final stage of FGD, only this token is connected to the classification MLP so that the classification information can be encoded to it during the forward process.

Positional encoding is a process that allows the model to observe the location information. In the positional encoding, a standard learnable design $E_{pos} \in \mathbf{R}^{(N+1)\times D}$ is followed in which a $D$ dimensional parameter is randomly initiated. Through the backpropagation process, the positional information can be encoded into the data processing workflow.

In the embedding process of the embedding module, feature patches are flattened and transposed into tokens $\left[x_p^1E; x_p^2E; ...; x_p^NE\right]$. Then, the class token $x_{class}$ is concatenated as the first input token, and the one-dimensional learnable positional encoding is added. With these designs, the feature maps of the stage $l$ are transformed into $N = N+1$ tokens, and each token has the exact dimension $D$. The output of the hybrid embedding process is $z_0$.

$$z_0 = \left[x_{class}; x_p^1E; x_p^2E; ...; x_p^NE\right] + E_{pos} \tag{1}$$

### 2) Focus block

In each backbone stage $l \in [1,2,3,4]$, the focus block transfers the early feature map $x_l \in \mathbf{R}^{E\times E\times C_l}$ (with the edge size of $E$ and the channel size of $C_l$) into the deeper stages as attention guidance which delivers the early-stage features such as textures. As shown in Fig. 1(b), the focus block of the FGD structure is composed of 3 main steps: (a) Attention module for processing the CNN features (b) Dual pooling path for gathering the attention biases inside the feature maps of each stage (c) Embedding module for transforming the attention biases into feature sequence of the Transformer decoder.

Firstly, the attention module used in the FGD focus block maintains the size $x_l^a \in \mathbf{R}^{E\times E\times C_l}$ of the input feature maps while processing the attention features from different CNN backbone stages. To obtain the attention information from the backbone, the SimAM block [38] is specially applied, the spatial biases of these early-stage neurons are captured by their activation. As a parameter-free attention module, the SimAM module is designed to identify distinguishing neurons in the Deep Neural Network based on findings from neuroscience and therefore enhances the feature maps by attention mechanism.

Secondly, a dual attention-gathering strategy is proposed to improve the global modeling process at the deep stages. Specifically, to obtain general and prominent features, feature maps $x_l^a \in \mathbf{R}^{C_l\times E\times E}$ after the attention module are considered by the design of 2 parallel pooling paths, including the max pooling and average pooling in (2) and (3) as follows:

$$x_l^{Maxpool} = MaxPooling(Attention(x_l)) \tag{2}$$

$$x_l^{Avgpool} = AvgPooling(Attention(x_l)) \tag{3}$$

Inspired by the gaze and glance process in human eyes, both pooling strategies share the same pooling window size $P$ and transform the feature maps into the feature patches with the edge size $p$. The outputs containing spatial attention information can be represented by $x_l^{Maxpool} \in \mathbf{R}^{C_l\times(E/P)\times(E/P)}$ and $x_l^{Avgpool} \in \mathbf{R}^{C_l\times(E/P)\times(E/P)}$.

Thirdly, the two separate 1x1 convolution layers share the same input and output channel sizes of $C_l$ and $D$, which are applied to alter the dimension of the feature maps, as shown in (4) and (5).

$$f_l^{Maxpool} = CNN1(x_l^{Maxpool}) \tag{4}$$

$$f_l^{Avgpool} = CNN2(x_l^{Avgpool}) \tag{5}$$

After the 2 CNN layers, feature patches $f_l^{Maxpool} \in \mathbf{R}^{D\times(E/P)\times(E/P)}$ and $f_l^{Avgpool} \in \mathbf{R}^{D\times(E/P)\times(E/P)}$ are transformed into $f_l^q \in \mathbf{R}^{(E/P)^2\times D}$ and $f_l^k \in \mathbf{R}^{(E/P)^2\times D}$ by flattening and transposing action, as shown in (6) and (7).

$$f_l^q = Transpose(Flatten(f_l^{Maxpool})) \tag{6}$$

$$f_l^k = Transpose(Flatten(f_l^{Avgpool})) \tag{7}$$

Lastly, the same embedding strategy in Fig. 2 is used inside the FGD focus block of each stage $l$ to obtain embedded guidance patches $q_l \in \mathbf{R}^{(N+1)\times D}$ and $k_l \in \mathbf{R}^{(N+1)\times D}$.

$$q_l = Concatenate(x_{class}, f_l^q) + E_{pos} \tag{8}$$

$$k_l = Concatenate(x_{class}, f_l^k) + E_{pos} \tag{9}$$

### 3) Decoder module

#### ① FGD decoder

To perform better global modeling and encode the attention information from different stages of CNN, the 4-decoder stacking structure is designed as the primary branch in the FGD structure, as shown in Fig. 1(a) and (c). The input of the first decoder is the feature patches (size of 145 * 768) processed by the hybrid embedding block, which is connected to the last stage of the CNN backbone in (10). The outputs of each decoder have the same size of 145*768, and the information flow is then transmitted to the next decoder stage. Each FGD decoder is stacked with the multi-head self-attention (MHSA), Multi-head Guided Attention (MHGA), and Feed-Forward Network (FFN) blocks. With the pre-norm strategy and the residual connection, the decoders can perform more robustly under different conditions.

In the workflow of each decoder, as shown in Fig. 1(c), a MHSA block (11) is firstly used to process the information from the last CNN block. Its purpose is to gather global information and achieve long-distance modeling by its self-attention structure. After the MHSA block, the FFN block (12) is used to stabilize the processing workflow. Because of the design of a 2-layer MLP connecting by non-linear activation units, FFN can also support the desire for generalization ability.

The focus block takes advantage of the inductive biases from different stages of the CNN backbone and encodes them as attention guidance. These features can be fused with the features of global modeling by the MHGA attention mechanism in (13). Finally, after the MHGA block, we use residual connection and FFN block (14) to stabilize the training process in the same way as the original Transformer decoder design [23]. The pre-norm strategy is applied to stabilize the gradient during the training process, therefore LayerNorm (LN) module is inserted before the MHSA, MHGA, and FFN blocks.

The decoders are connected to each other by the stacking design at the last stage of the FGD structure in (15), only the class token $x_{class}^{output}$ of the output sequences is connected to the MLP classification head.

$$z_0 = hybrid_Embedding(CNN_backbone(x)) \tag{10}$$

$$z_l^1 = MHSA(LN(Z_{(l-1)})) + z_{(l-1)}, l = 1,2,3,4 \tag{11}$$

$$z_l^2 = MLP(LN(z_l^1)) + z_l^1, l = 1,2,3,4 \tag{12}$$

$$z_l^3 = MHGA(\mathrm{LN}(z_l^2), f_l^q, f_l^k) + z_l^2, l = 1,2,3,4 \tag{13}$$

$$z_l = MLP(LN(z_l^3)) + z_l^3, l = 1,2,3,4 \tag{14}$$

$$z_4 = \left[x_{class}^{output}; x_{output}^1; x_{output}^2; \ldots; x_{output}^N\right], x_{output} \in \mathbf{R}^{1\times D} \tag{15}$$

### ② Multi-head Self-attention (MHSA)

The Transformer models are known for the long-distance modeling ability of MHSA [23], which contributes to its global modeling process. As shown in Fig. 3, MHSA has the tremendous advantages of capturing embedded feature maps which achieve its strength in the global modeling process of the FGD structure.

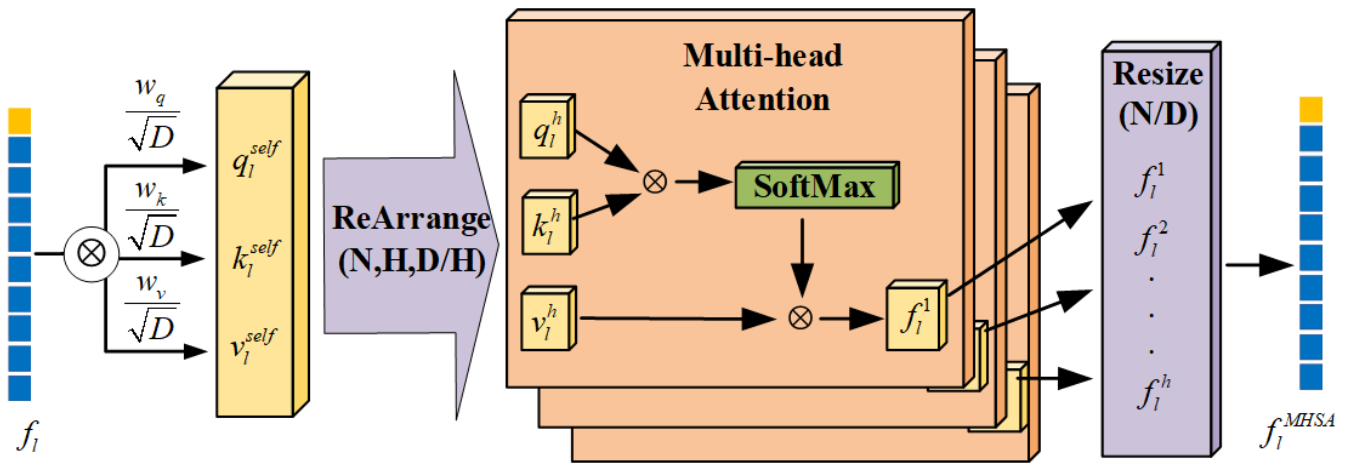

Fig. 3. The architecture of the multi-head self-attention.

In the MHSA block, the given feature patches $f_l$ of the stage $l$ are transformed into three sequences called $q_l^{self}$, $k_l^{self}$, $v_l^{self}$ by the scale inner dot product with three learnable matrices $w_q$, $w_k$, $w_v$ as $q_l^{self} = w_q \bullet f_l / \sqrt{D}$, $k_l^{self} = w_k \bullet f_l / \sqrt{D}$, $v_l^{self} = w_v \bullet f_l / \sqrt{D}$.

After the transform, the resize calculation is applied to partition them into $\left[q_l^1, q_l^2, \ldots, q_l^h\right]$, $\left[k_l^1, k_l^2, \ldots, k_l^h\right]$ and $\left[v_l^1, v_l^2, \ldots, v_l^h\right]$ where $h$ is the number of heads. In each head $h$, for a given embedded patch $f_l$, $f_l^h = SoftMax((q_l^h)^T \bullet k_l^h)^T \bullet v_l^h$. The output $\left[f_l^1, \ldots, f_l^h\right]$ is resized to reduce the dimension and the $f_l^{MHSA} \in \mathbf{R}^{N\times D}$ can be obtained.

### ③ Multi-head Guided-attention (MHGA)

The attention information $q_l$ and $k_l$ captured by the dual information flow in the FGD focus block play a significant role in MHGA. As shown in Fig. 4, different from the MHSA, the MHGA block is aiming for processing attention information from the FGD focus blocks and using it to encode additional spatial information biases within the global modeling process of the FGD decoder blocks.

The inputs $q_l$ and $k_l$ which contain prominent and general attention information from early CNN stages are transformed into feature patches $q_l^{guide} = w_k \bullet f_l^q / \sqrt{D}$ and $k_l^{guide} = w_k \bullet f_l^k / \sqrt{D}$. Similar to MHSA, the scaled inner dot product and SoftMax layers are designed after the resizing operation to split the feature patches into different heads. Within each head, $v_l^h$ is multiplied with $SoftMax((q_l^h)^T \bullet k_l^h)^T$ to obtain $G_l^h$. The multi-head information is combined by rearranging and resizing operations for the output $f_l^{MHGA}$.

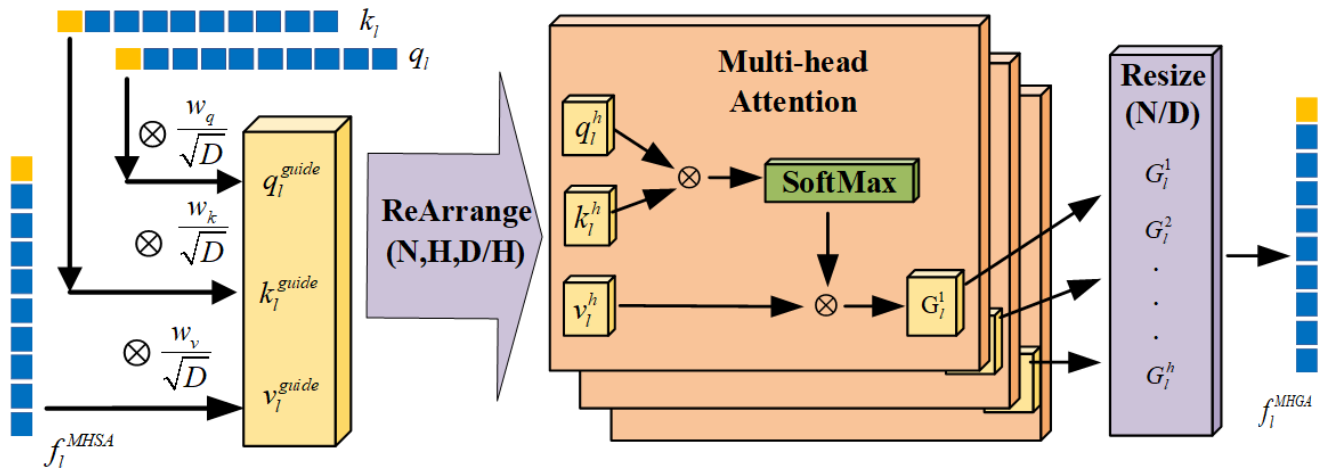

Fig. 4. The architecture of the multi-head guided-attention.

## C. Multi-stage Hybrid Transformer

The proposed MSHT model is a hybrid model that integrates CNN with Transformer, where the main feature extraction workflow and the intermediate outputs of the backbone CNN are considered. The feature maps are downsampled through the backbone CNN stages with the size of 256*96*96, 512*48*48, 1024*24*24, and 2048*12*12 from the first to the fourth stage.

After the processing of the CNN backbone, the hybrid embedding block is introduced to transform the feature maps from the last CNN stage into the input feature patches of the Transformer blocks, sizing from 2048*12*12 to 144*768. At the end of the embedding block, the size of the output feature patches is expanded from 144*768 to 145*768 due to the concatenation of an empty $x_{class}$ with the size of 1*768.

Combining with the CNN structure, each focus block firstly downsamples the feature maps from different stages to the same edge size as the last feature map (12*12) and then embed the attention-guided information into two patch sequences with the size of 145*768. As shown in (11-14), the global modeling process by the Transformer decoders can model the embedded sequences with attention guidance sequences. After the four stages of the FGD structure, the first dimension $x_{class}^{output}$ which carries the classification information in the output sequences, is connected to the MLP classification head.

Lastly, the MLP combined with the SoftMax layer projects the class token $x_{class}^{output}$ (size of 1*768) to the class numbers which represent the predicted confidence of each class.

# III. D. Experiment and Results

## A. Dataset

The ROSE diagnosis was implemented in Peking Union Medical College Hospital (PUMCH), and the data were collected under the supervision of senior pathologists. The pancreatic cytopathological images were obtained from EUS-FNA with ROSE examinations and 4240 images in total were sampled after the diff-quik giemsa staining procedure. The enrolled images were performed by two microscope digital cameras (Basler ScA1 and Olympus DP73) with Olympus BX53 and Nikon Eclipse Ci-S microscopes. From 2019 to 2021, a total of 1518 pancreatic cancer images and 2722 normal pancreatic cell images were collected. During the data sampling process, the images were saved in 'jpg' format with resolutions of 1390*1038, 2400*1800, and 5480*3648 under the same magnification of 400 times. Since all input images should share the same length-to-width ratio, they were resized to 1390*1038 to maintain the same magnification factor in the range. The classification labels were confirmed by the senior pathologists of PUMCH.

## B. Experimental Setting and Hyperparameter Setting

To evaluate the models on the ROSE images, we adopted a 5-fold training setting. Firstly, we randomly divided the 4240 images into two groups: a training-validation set and an independent test set with a ratio of 8:2. Then the training-validation set was randomly divided into five datasets, representing 5-folds with approximately the same number of images in each fold. In the fold $k$ ($k \in [1,2 \dots 5]$), the $k$ fold dataset was used as the validation dataset, while the data from the remaining four folds were used as the training datasets. In each experiment, the model was trained five times individually, with different validation and training datasets each time, but the independent test dataset was shared. In all folds of the experiments, 50 epochs were trained and tested on the training and validation datasets. The model for a given epoch, which showed the best accuracy on the validation dataset, was saved as the output model of the fold $k$.

A certain data-augmentation strategy was implied in the experiments to recreate views under the microscope, as shown in Fig. S1. During each training process, the input images were randomly rotated, and a center area with a size of 700*700 pixels was cropped as the input data. Random horizontal flip and random 'color-jitter' (including brightness, contrast, saturation, and HUE shifting operations) were used to recreate the white balance shifting and other perturbations during the sampling process. In each validating and testing process, only the 'CenterCrop' operation was used to reserve the central pixels inside the 700*700 boundary. The Adam optimizer was used with a learning rate of 6e-5 and a momentum of 0.05. In the training process, the cosine learning rate decay strategy was adopted to reduce the learning rate ten times sequentially. The counterparts of the MHST model were trained with the same hyperparameter setting or better performed hyperparameters.

The experiments were carried out and recorded online on the Google CoLab pro+ platform. In each experiment, a 16 GB Nvidia P100-PCIe GPU was offered with Python version 3.7.12 and Pytorch version 1.9.0+cu111. The model was built based on the Pytorch [39] and timm [40] library. The implementation details along with the entire experiment scripts and records have been released online.

## C. Evaluation Criteria

The experimental results of MHST and its counterparts were measured by Accuracy (Acc), Precision (Pre), Recall (Rec), Sensitivity (Sen), Specificity (Spe), Positive predictive value (PPV), Negative predictive value (NPV), and F1-score. During the measurement of the 2-class classification task (positive or negative), the criteria were calculated by the True Positive (TP), True Negative (TN), False Positive (FP), and False Negative (FN), all of the detailed results were recorded to indicate their research values. To reveal the interpretability of the models, the classification-activating-mapping (CAM [41]) method was used to visualize the attention regions.

## D. Results of MSHT

In Table I and Table SI, several evaluation criteria were used to assess the experimental result, and a 5-fold average result for each indicator was presented to indicate the overall performance of MSHT and its counterparts.

In the 5-fold training process, the MSHT achieves an average of 97.53%, 98.07%, 96.56%, 96.54%, and 98.08% for Acc, Spe, Sen, PPV, and NPV. Meanwhile, the average performance of these indicators during the validating process are 94.37%, 96.69%, 90.20%, 93.93%, and 94.67%. In the independent test dataset, 95.68 % of images are correctly classified, and MSHT achieved 96.95%, 93.40%, 94.54%, and 96.35% for Spe, Sen, PPV, and NPV. The results of MSHT are significantly better than the early work by Hashimoto et al. (with 93% and 89% accuracy [21][22]), and MSHT is fully validated on a larger dataset of 4240 images compared with their 1440 images.

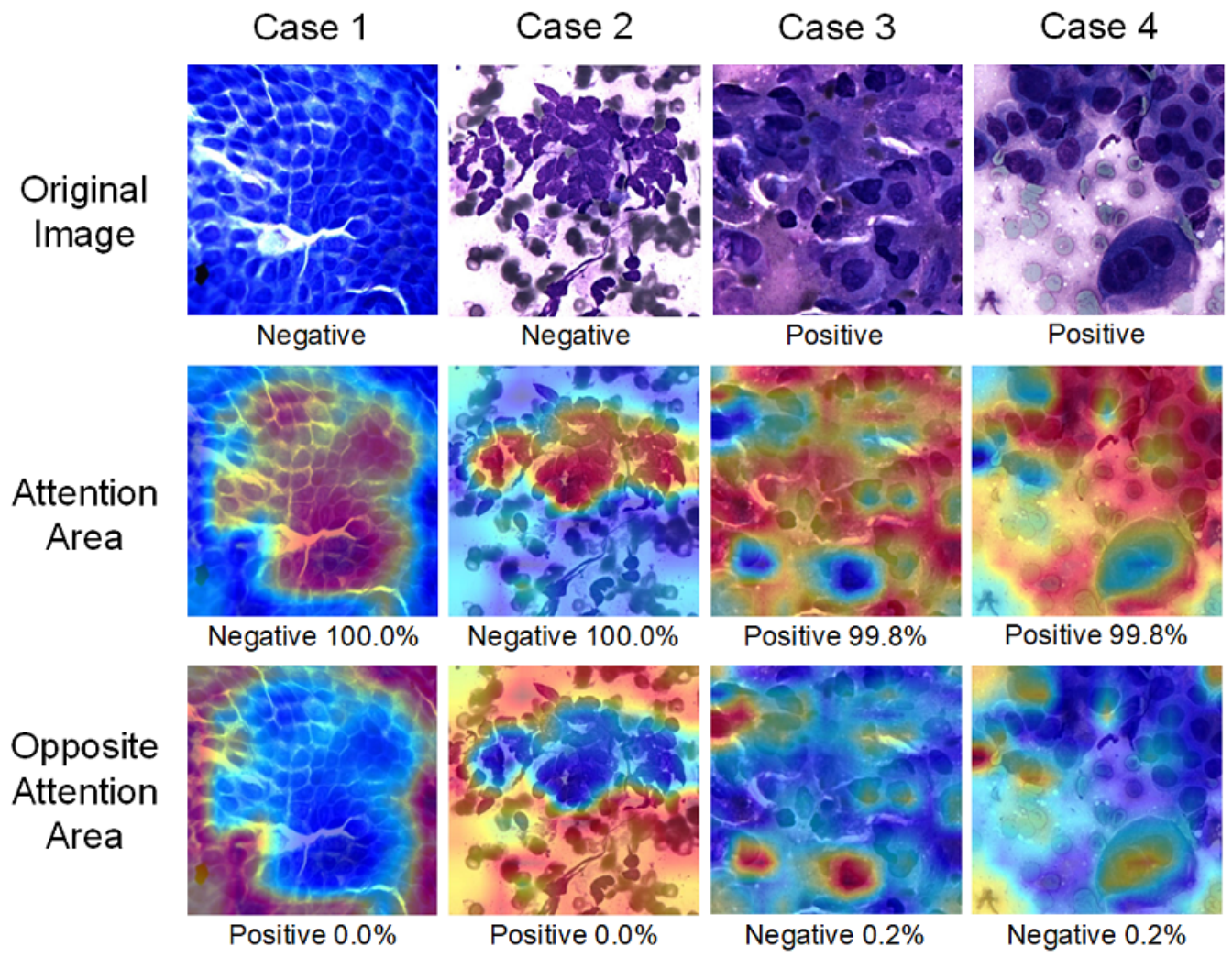

Fig. 5. Examples of typical samples and the attention regions of MSHT.

The results indicate that the MSHT has achieved encouraging results on the ROSE images, as shown in Fig. 5 and Fig. S2, presenting solid interpretability through the visualization of its attention regions by the Grad-CAM technique. In Fig. 5, four typical cases are presented with predictive confidence, and the decision boundaries of the feature regions are represented by heatmaps. Within each heatmap generated by the CAM analysis, the red areas indicate where the certain classification output is highly correlated, while the blue areas are correlated with other categories. In case 1, two typical negative samples are shown with clear background and complex surroundings. The attention regions indicate the MSHT has a preference on each cell area and it has clearly captured the long-distance distribution. In case 2, MSHT has evidently identified the pancreatic cells without the redundant attention spent, even in the noisy surroundings. In cases 3 and 4, MSHT clearly identifies cancerous cells despite the fact that there were fewer positive samples than negative samples due to the imbalance of the dataset. The spatial difference of cancer cells and their distinctive distribution are correctly captured by the MSHT, as the attention regions indicate. The accurate attention regions prove the robust interpretation of the proposed MSHT and imply its clinical potential.

TABLE I
RESULTS OF THE COUNTERPART MODELS ON THE ROSE DATASET

| *Model* | *Acc (%)* | *Spe (%)* | *Sen (%)* | *PPV (%)* | *NPV (%)* | *F1_score (%)* |
|---|---|---|---|---|---|---|
| *ResNet50[28]* | 95.02 | 95.51 | 94.13 | 92.17 | 96.70 | 93.12 |
| *VGG-16[26]* | 94.92 | 95.66 | 93.60 | 92.42 | 96.44 | 92.95 |
| *VGG-19[26]* | 94.83 | 96.03 | 92.67 | 93.02 | 95.96 | 92.78 |
| *Efficientnet_b3[32]* | 93.29 | 95.48 | 89.37 | 91.80 | 94.19 | 90.51 |
| *Inception V3[29]* | 93.84 | 94.49 | 92.67 | 90.35 | 95.86 | 91.49 |
| *Xception[31]* | 94.69 | 96.07 | 92.21 | 92.91 | 95.68 | 92.55 |
| *Mobilenet V3[33]* | 93.44 | 95.11 | 90.43 | 91.20 | 94.70 | 90.80 |
| *ViT[24]* | 94.50 | 95.26 | 93.14 | 91.63 | 96.14 | 92.37 |
| *DeiT[44]* | 94.52 | 95.04 | 93.60 | 91.34 | 96.41 | 92.42 |
| *Swin Transformer[45]* | 94.92 | 95.18 | **94.46** | 91.74 | **96.87** | 93.03 |
| ***MSHT(Ours)*** | **95.68** | **96.95** | 93.40 | **94.54** | 96.35 | **93.94** |

Additionally, the problem of misclassification that needs to be overcome is given by taking two examples, as shown in Fig. 6, Fig. S3 and Fig. S4. A specific image has been misclassified to positive condition by three of the five 5-fold MSHT models. By the analysis of senior pathologists, the reason lies in the fluctuation of the squeezed sample, which misleads MSHT by the shape of the cells. A few positive samples are misclassified as negative ones. Compared with senior pathologists, the small number of the cells makes it difficult for MSHT to distinguish cancer cells by their arrangement and relative size information.

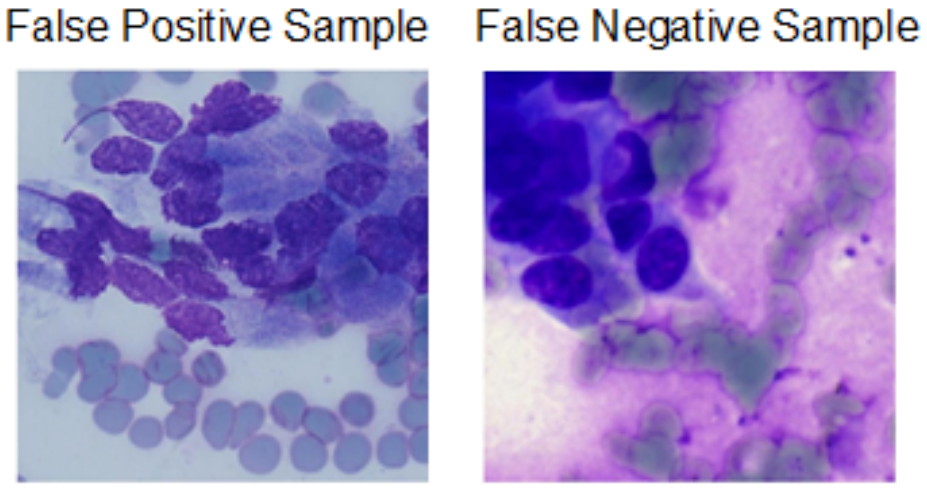

Fig. 6. Examples of misclassified samples.

## E. Results of Comparison Models

Due to the limited work of this field, we evaluated the proposed MSHT with seven widely applied state-of-the-art CNNs including: ResNet50 [28] (2016), VGG-16, VGG-19 [26] (2014), EfficientNet_b3 [32] (2019), Inception-V3 [29] (2016), Xception [31] (2017) and MobileNet-V3 [33] (2019).

As the first work to introduce the Transformer into ROSE image analysis, three cutting-edge Transformer-based models (vision transformers) from the computer vision field were used as the comparison models responsibly, including ViT [24] (2020), DeiT [44] (2020), and Swin Transformer [45] (2021). It should be noted that Transfer learning was used for all models with the official weight of models pre-trained on the ImageNet [46]. The models were compared with the same criteria on the test dataset after they all converged at the same

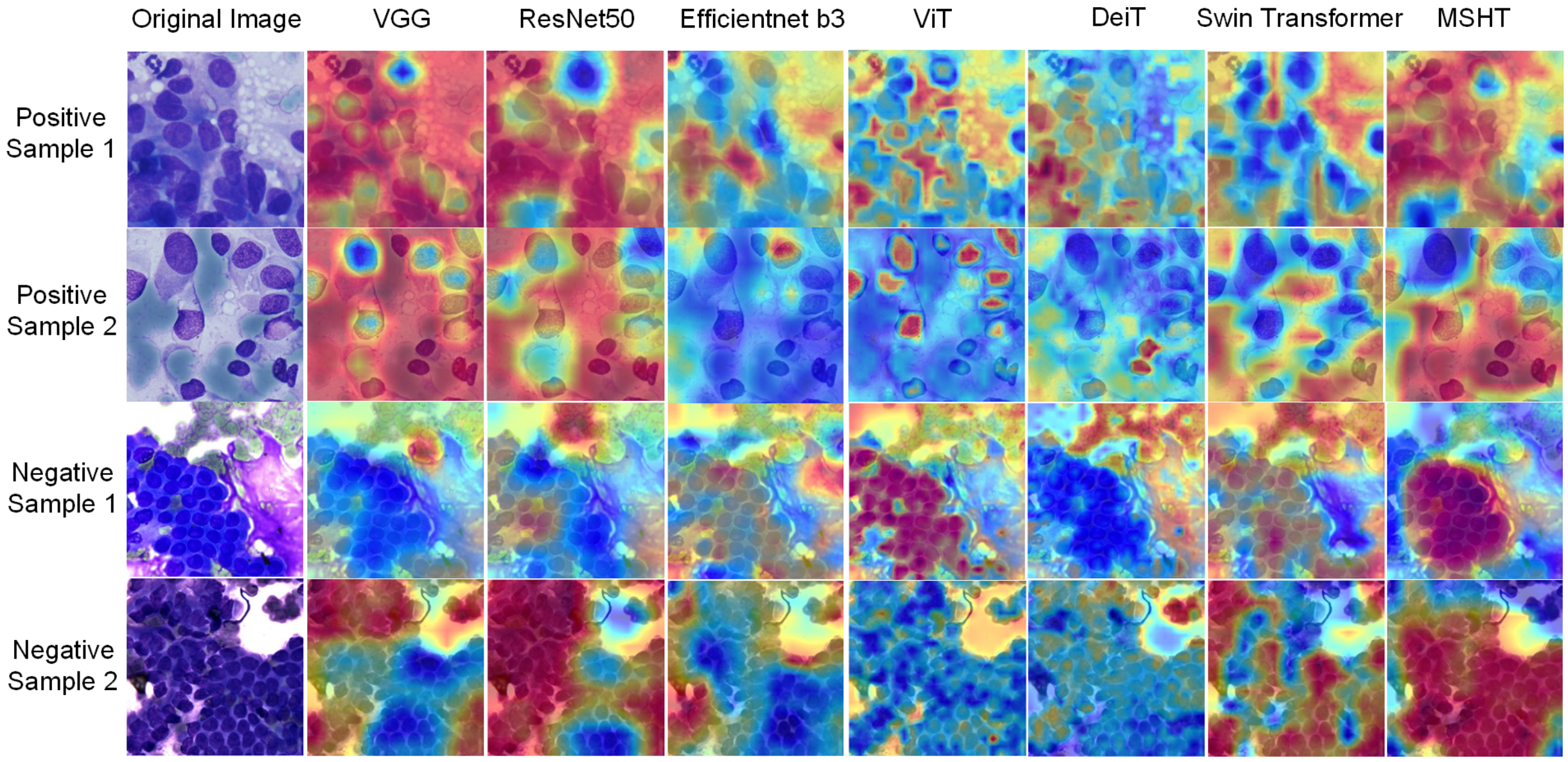

Fig. 7. The attention regions of different models in ROSE image analysis.

hyperparameter setting.

Table I shows the 5-fold average results of models on the testing process of ROSE image classification, which indicates that the proposed MSHT achieves the highest Acc overall (95.68%) and outperforms its counterparts significantly. In the experiments, most of the CNNs and Transformers except ResNet50 achieve approximate Acc less than 95%, while only ResNet achieved 95.02%, slightly higher than 95%. Although the imbalance of the dataset (1518 positive images vs. 2722 negative images) increases the difficulty for models to recognize positive samples, compared with other models, MSHT has achieved significantly higher results in Spe and PPV which contribute to the higher overall performance in Acc and F1-scores.

In Table I, two CNNs and two Transformers achieve higher Sen and NPV than MSHT. The optimized VGG-16 and DeiT models achieve slightly higher Sen (+0.20%, +0.20%) and NPV (+0.09%, +0.06%) compared with MSHT. The best-performed CNN counterpart ResNet50 achieves 0.73% higher Sen and 0.34% higher NPV than MSHT, but it shows obvious lower Spe (-1.43%) and PPV (-2.37%) results. Furthermore, the best-performed Transformer model Swin Transformer achieves higher in Sen (+1.06%) and NPV (+0.52%) with much lower results in Spe (-1.77%) and PPV (-2.80%). Since the negative samples are more than the positive ones, higher Sen and NPV are easier to obtain than other criteria. Such results indicate that the high Acc performed counterparts share the same biases on the negative samples. The MSHT outperforms the rest six counterparts by higher results in all six criteria Acc, Spe, Sen, PPV, NPV, and F1-score. The results indicate that the proposed MSHT has better robustness under unbalanced data conditions, while many counterparts are influenced. In the AI-aided diagnosis process for ROSE images, Spe and PPV are more important than the opposite indicators Sen and NPV. This is because the misdiagnosis of AI-system leads to the inadequate sampling of pancreatic tissues and further results in reducing the final diagnostic accuracy, while the missed diagnosis usually leads to more sampling of pancreatic tissues and does not affect the final diagnostic accuracy.

In terms of interpretability, as shown in Fig. 7, the models perform differently when their attention regions are visualized by the Grad-CAM technique. The CNN models show acceptable attention regions when dealing with negative samples, while they focus on both cells and background when facing positive samples. Under noisy conditions, as shown in negative sample 1, the CNNs tend to suffer from background perturbation, which is not ideal in clinical applications. In contrast, Vision Transformers show different biases. Briefly, the Transformers can capture the features of the pancreatic cells as they have clear attention regions on them. In negative sample 2, the attention regions are discrete while the cells are close to each other, which indicates the limitation of spatial related modeling in the Transformers. The Transformers easily focus on the background of the positive samples, indicating that the models have not learned the correct patterns due to the limited samples.

The MSHT outperforms CNNs as its attention regions cover the distinctive cells in most cases. Furthermore, the MSHT shows its robust focus in noisy situations. Compared with the Transformers, in addition to being able to focus on cells globally, its attention regions are more aggregated following the trend of cells. In most cases, as shown in Fig. 7, MSHT can correctly distinguish the samples and focus on the cell groups like the senior pathologists.

### F. *Results of MSHT Ablation Studies*

To evaluate the structural improvement and explore the efficiency of the proposed MSHT, a series of ablation studies were drawn out under the same training and validating setting. Except for the pre-training experiment, MSHT ablation counterparts had their backbone weights initiated by the official ResNet50, and the weights of the remaining structures were initialized randomly for a fair comparison.

#### 1) *The effectiveness of MSHT FGD structure*

Firstly, to evaluate the effectiveness of the FGD structure, we designed the Hybrid1, Hybrid3 models. In the Hybrid1 model, four stages of ResNet50 [28] structure and eight Transformer encoder modules were implemented, in which the same calculation mainstream as the proposed MSHT was recreated by a stacking design. Meanwhile, the 3-stage design improves the resolution of the CNN feature maps but requires more calculation due to its feature map expansion (edge size increased from 12 to 24). Therefore, we designed the Hybrid3 to compare the proposed FGD structure in terms of the stage depth and the size of the feature maps.

As shown in Table II, the proposed 4-stage design of the MSHT (Hybrid2_No_PreTrain) model achieves the best performance in terms of Acc, Sen, NPV, and F1-score with 95.30%, 93.66%, 96.47%, and 93.45%. Meanwhile, the Hybrid1 model achieves the same result in Sen, but it only achieves 94.90% ACC and 92.93 F1-score in general criteria. The results indicate there is a distinctive gap in the absence of the FGD structure. The stacking design may limit the pattern modeling ability in the deeper layers, as the mainstreams of the model are the same in the Hybrid1 and MSHT (Hybrid2_No_PreTrain) design.

TABLE II
ABLATION STUDIES

| **Groupe** | *Model* | *Acc (%)* | *Spe (%)* | *Sen (%)* | *PPV (%)* | *NPV (%)* | *F1_score (%)* |
|---|---|---|---|---|---|---|---|
| *Connecting type* | *Hybrid1* | 94.90 | 95.59 | **93.66** | 92.24 | 96.45 | 92.93 |
| | *Hybrid3* | 94.73 | **96.54** | 91.49 | **93.66** | 95.33 | 92.55 |
| *Sharing-parameter design* | *Hybrid2_No_CLS_Token* | 94.85 | 96.25 | 92.34 | 93.24 | 95.77 | 92.77 |
| | *Hybrid2_No_Pos_emb* | 94.71 | 96.10 | 92.21 | 93.00 | 95.71 | 92.56 |
| *Change attention blocks* | *Hybrid2_No_ATT* | 94.52 | 95.44 | 92.87 | 91.96 | 96.02 | 92.38 |
| | *Hybrid2_SE_ATT* | 94.71 | **96.25** | 91.95 | 93.25 | 95.56 | 92.56 |
| | *Hybrid2_CBAM_ATT* | 95.11 | 95.96 | 93.60 | 92.84 | 96.42 | 93.20 |
| *Pre-training transfer learning* | *Hybrid2_No_PreTrain* | **95.30** | 96.21 | **93.66** | 93.28 | **96.47** | **93.45** |
| | ***MSHT*** | **95.68** | **96.95** | 93.40 | **94.54** | 96.35 | **93.94** |

Taking account of the complexity of FGD, the standard 4-stages of ResNet structure has smaller feature sizes and higher dimensions carrying deeper CNN features, which is used in MSHT. The 3-stage Hybrid3 model reaches the highest Spe and PPV with 96.54% (+0.33%) and 93.66% (+0.38%), which are slightly better than the 4-stages design model. Since the Hybrid3 model only reaches 94.73% Acc (-0.57%) with more calculation cost in the experiments, it can be seen that the number of stages of FGD plays a crucial role. These results indicate that the FGD structure is effective and that the 4-stage design achieves the best results cost-efficiently.

#### *2) The effectiveness of class token and positional encoding*

Following the computer vision studies, the class token and positional encoding are pivotal to revealing the performance of the Transformer based model. In MSHT, the global sharing class token and positional encoding are designed to work as the messenger throughout the FGD structure. Therefore, the Hybrid2_No_CLS_Token and Hybrid2_No_Pos_emb are designed specially, without using the class token and positional encoding in the FGD structure.

As shown in Table II, the results indicate the MSHT design achieves evidently higher performance in all six criteria. Without the class token, the model Hybrid2_No_CLS_Token reaches 94.85% ACC and 92.77% F1-score, which are slightly higher than the performance of Hybrid2_No_Pos_emb with 94.71% and 92.56% only. The results prove that class token design can influence the outcome but the positional encoding is more critical to the model, as the knowledge of the distribution of the cells contributes to the global modeling process of the Transformer.

#### *3) The effectiveness of attention module in focus block*

In the FGD focus block, an attention module is employed before the dual pooling layers to stabilize the training process. The focus block equipped with the widely used CBAM [42] module and SE [43] module are compared as Hybrid2_CBAM and Hybrid2_SE. Noticeably, the attention module SimAM deployed in the focus block is a parameter-free structure that reduces the model size and complexity, the Hybrid2_No_ATT is therefore designed to prove its effectiveness.

The proposed MSHT with SimAM module achieves a higher result in all six criteria compared with no attention module design and CBAM attention module version. Compared with no attention design Hybrid2_No_ATT, un-pre-trained MSHT (Hybrid2_No_PreTrain) achieves 0.78% higher Acc and 1.07% higher F1-score. Only a slightly higher difference can be shown compared with Hybrid2_CBAM in general criteria, Hybrid2_No_PreTrain achieves 0.19% and 0.25% higher in ACC and F1-score. Only in a specific indicator Spe, the Hybrid2_SE achieves slightly higher (by +0.04%) than Hybrid2_No_PreTrain. The proposed MSHT achieves the ambition of feature converting in FGD with a cost-efficient module SimAM, which achieves the best general result at the lowest calculation cost.

## IV. Discussion and Conclusion

Through the stage-wise hybrid design, the proposed MSHT introduces the cutting-edge Transformer structure into ROSE image analysis and performs more sophisticated results than the backbone CNNs. Together with the CNN feature extraction structure, the introduction of the Transformer module improves the global modeling capabilities by globally integrating the local features of cells and the relationships between cells.

Compared with the state-of-the-art models, MSHT achieves significantly higher Acc and F1-score. At the feature extracting stages, the MSHT model inherits the strength of the inductive bias of the CNNs blocks, which contributes to robust performance under complex surroundings. Benefiting from the global modeling design of attention mechanism, by introducing the Transformer module, MSHT uses the attention information from each stage of CNN backbone to guide the global modeling process. Compared with the pure attention-mechanism-based models ViT and DeiT, MSHT performs higher overall results with the multi-stage hybrid structure converting attention guidance from the CNN backbone.

Using the early-stage feature maps of CNN, the attention biases are obtained by the FGD structure to guide the process of global modeling. Two different global sharing designs, class token and positional encoding carrying the through-out information are used in each embedding process, leading to a more significant improvement in the results than other ablation models. The FGD structure influences the effectiveness of the Transformer blocks and CNN blocks. Compared with the directly stacking strategy without FGD structure, feature maps of different CNN stages are used as the attention biases, which directly contribute to the modeling process of Transformer blocks. Focusing on the number of stages in the CNN backbone, deeper stage (four stages vs. three stages) design enhances the stability and robustness when the model is faced with small medical datasets.

In medical image analysis, data scarcity is pivotal in performing its full potential. To alleviate such limitation, the transfer learning strategy is hired following the same limitation and the pre-trained model obtains better results in the training, validation, and test datasets. The proposed model is pre-trained on the ImageNet-1k [46] for 150 epochs and with a batch size of 210 on 3 Nvidia A6000 GPUs. The learning rate is set to 1e-6 and with a cosine learning rate decay for ten times. In Table II, comparing with un-pre-trained MSHT model, the proposed MSHT achieves distinctive higher Acc (+0.38%), Spe (+0.74%), PPV (+1.26%), and F1-score (+0.49%). The overall criteria Acc and F1-score are distinguished higher, but the data-driven Sen and NPV are slightly lower than the randomly initialized counterparts.

In conclusion, the proposed MSHT model achieves the state-of-the-art classification performance on the cytopathological ROSE image analysis via its unique multi-stage hybrid architecture which can effectively combine the advantages of CNN and Transformer. Along with the clinical innovation ROSE technique, the MSHT model can help to explore the pathologist-free EUS-FNA procedures and expand its potential for wider application in the diagnosis of pancreatic cancer.

## Supplementary

In MSHT, we take advantage of spatial local attention information from different stages of CNN by taking different grained feature maps and using the attention module (inside the focus block) to transform them into attention guidance in FGD structure.

To process the global modeling with attention guidance, we design the FGD decoder based on the original Transformer decoder to capture the early attention information focusing on local features at different stages in the CNN structure. The decoder block is designed with a similar structure in the encoder of ViT, we reserved the first multi-head self-attention (MHSA) and feed-forward Network (FFN) blocks in Transformer design, but for the second MHSA like block, we used the multi-head guided attention (MHGA) instead. With this design, the FGD decoder block can process the feature maps from the previous block and take advantage of the attention biases of earlier CNN layers by using the attention-guided focus block and the guided- attention block. Focusing on global modeling, by fusing the deep stage patterns with robust CNN biases from different stages, we achieved our intention by introducing a limited calculation cost.

In the FGD focus block, the attention module of SimAM [38] is introduced. In its original paper, $e_t^*$ is the energy level of a certain neuron corresponding with the surrounding ones, the lower it is, the more prominent the neuron t would perform. In Equation, $\hat{\sigma}$ and $\hat{\mu}$ are calculated by the feature maps, $\lambda$ is a hyperparameter.

$$e_t^* = \frac{4(\hat{\sigma}^2 + \lambda)}{(t-\hat{\mu})^2 + 2\hat{\sigma}^2 + 2\lambda}$$

By taking the activations of the neurons into account, SimAM enhances the feature maps with the algorithm as follows:

**Algorithm 1** Simam Algorithm

**Input:** Feature maps X of size (batch_size, channel_size, width, height);

**Output:** Attention enhanced feature maps Y;

**for** a input image x in batch **do**
  Get spatial size $n = x.width * x.height - 1$;
  **for** a feature slice of size (width*height) **do**
    $d = (x - average(x))^2$;
  **end for**
  Calculate all the importance of the input image $E_{inv} = \frac{d}{4*(v+lambda)} + 0.5$;
  Attention activate output $y = x * sigmoid(E_{inv})$;
**end for**
Group the output at batch level, return the result as Y

As shown in Fig. S1, the intention of data augmentation is to recreate the clinical sampling condition, thus increasing the generalizability of MSHT. The random rotation and color-shifting methods are especially applied in the training process of all the experiments besides the widely applied 'HorizontalFlip'. By using PyTorch 'ColorJitter' with brightness = 0.15, contrast=0.3, saturation=0.3, and hue=0.06, each training sample will be projected to a general-domain-based image.

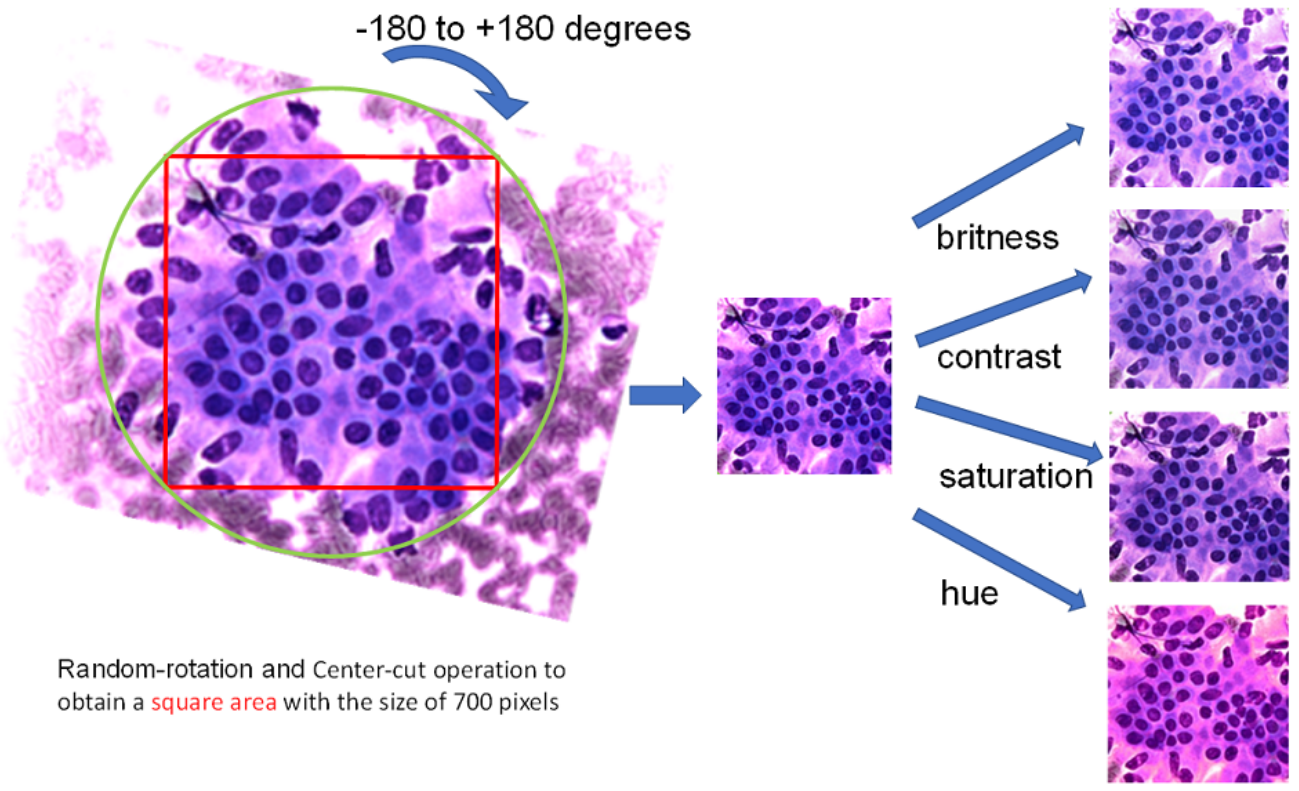

Fig. S1. Examples of data-augmentation methods recreating the clinical sampling views.

In Table SI eight widely applied CNNs and three state-of-the-art Transformer-based models are compared in the process of ROSE image analysis as counterparts to MSHT. Generally, the models achieved acceptable results, varying from 90.91 % to 95.02 % in Acc. The distributions of results indicate that the highest criteria show in the training dataset and the lower ones in the validation and the test datasets. Due to the limited dataset, the test dataset is twice than the validation dataset which leads to more confident results despite the perturbation of samples. Since the models are saved when they performed the highest marks in the validation dataset, the results in the training and validating process are not as supportive as the testing process. In the training and validation dataset, the MSHT achieves similar results compared with the highest performed models, but it achieves significantly higher results in the test dataset. The results prove that the MSHT has achieved distinctively higher performance.

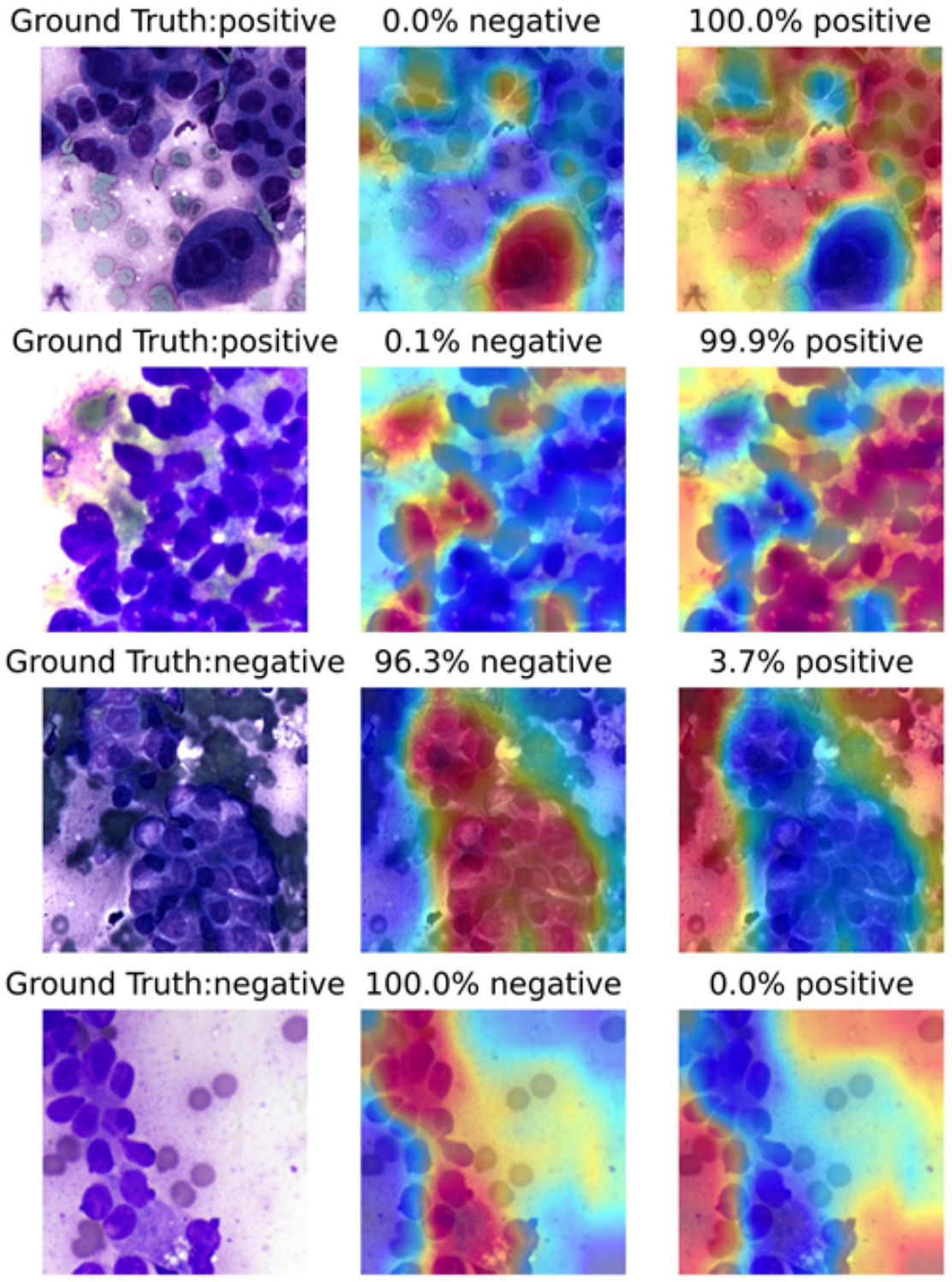

Fig. S2. Examples of correctly classified samples.

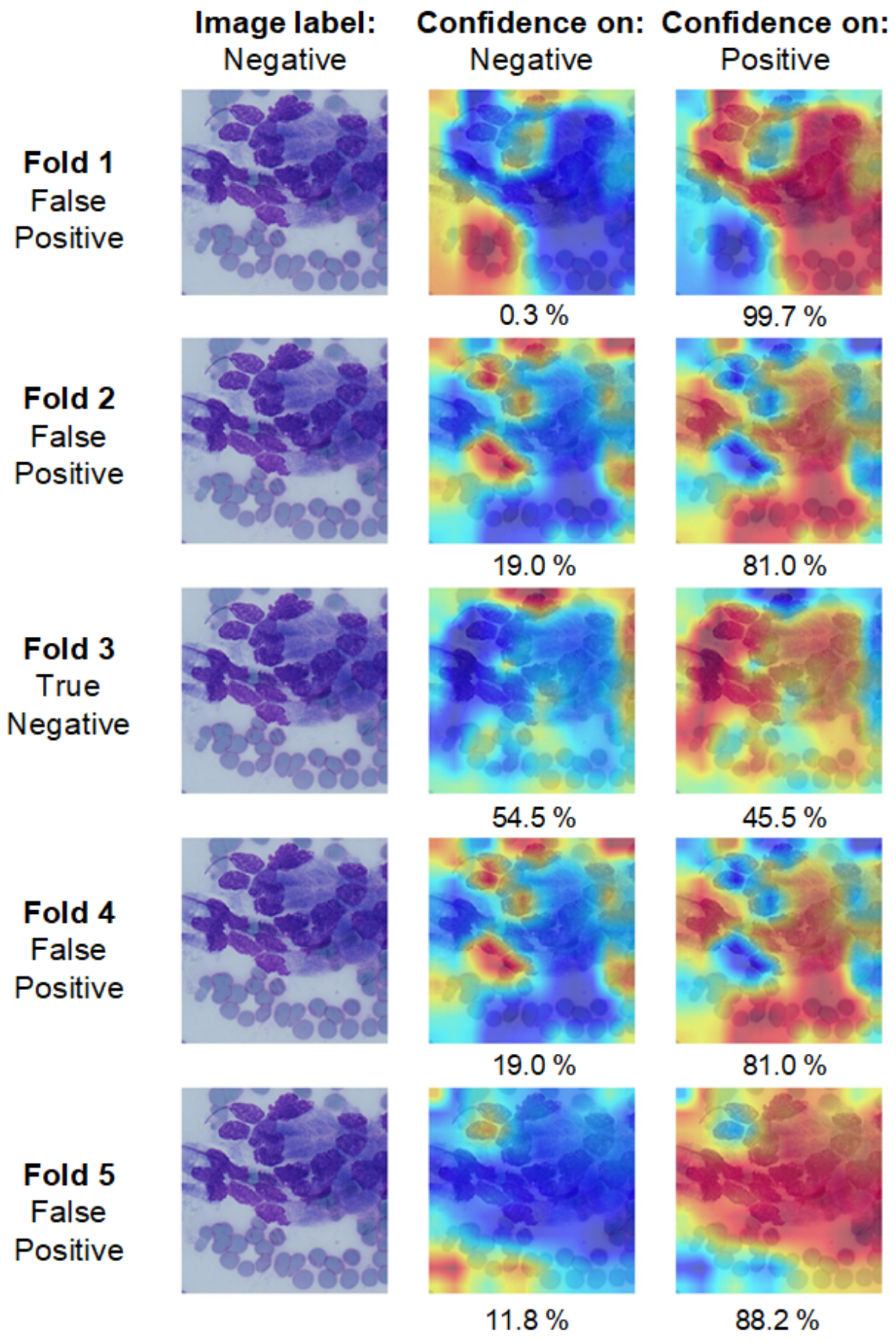

Fig. S3. Examples of misclassified (False positive) samples and the attention regions of the MSHT.

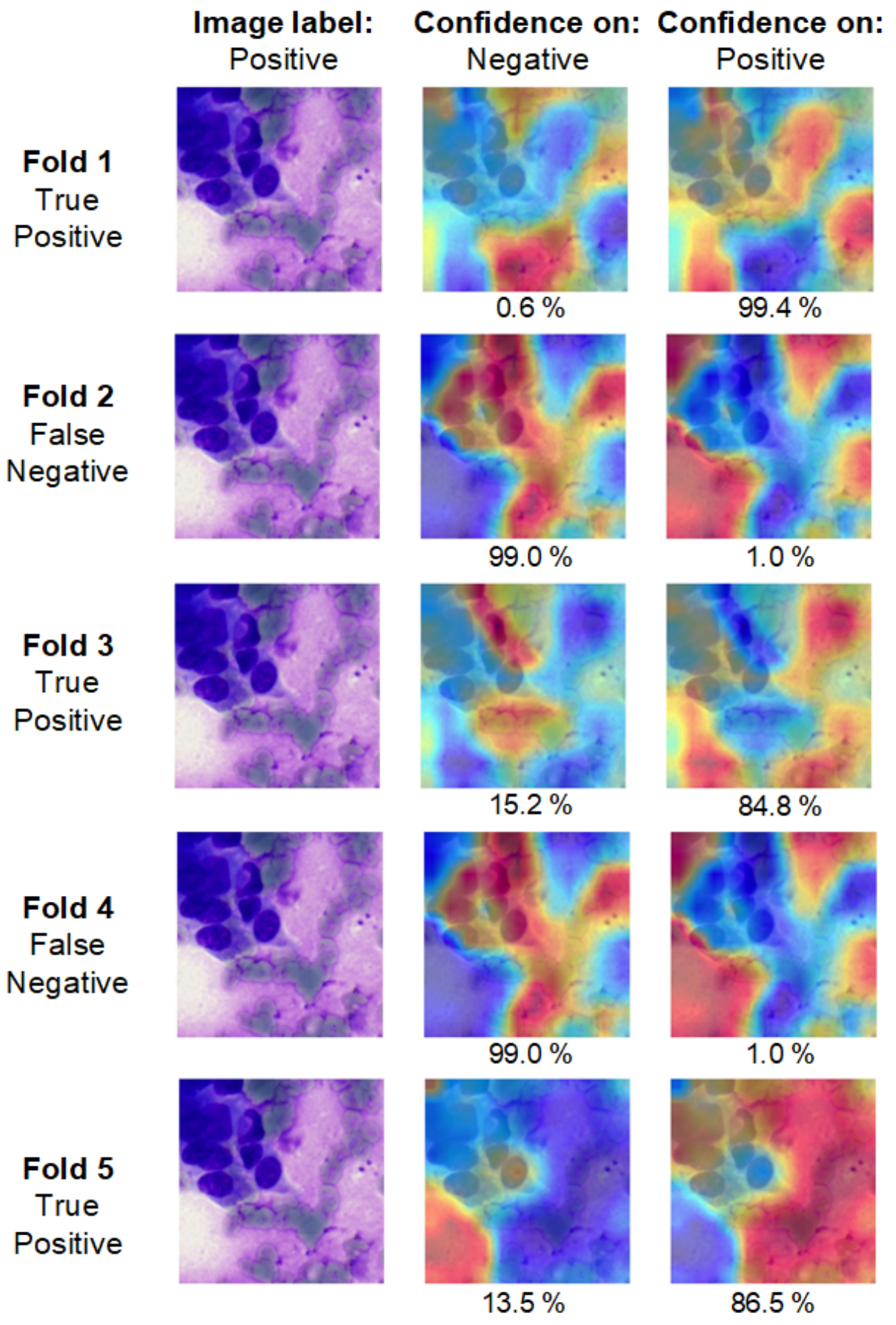

Fig. S4. Examples of misclassified (False Negative) samples and the attention regions of the MSHT.

The MSHT performs delightfully in the study of interpretability, as shown in Fig. S2. Most cases can be identified correctly, even when the surrounding of cells is complex and confuses junior doctors. Moreover, the classification confidence is decisive in most cases, which indicates its performance is reliable.

As mentioned in the article, for some instances, the decision boundaries are unclear as the interpreted study reveals pivotal information, as shown in Fig. S3. A particular negative sample is misclassified to its counterpart due to the distorted perturbation in the sampling or staining process. The models in 4 of all five folds identify it as the positive, and only a model of 5-folds experiments distinguish it correctly but with only vague decision boundaries (54.5 % vs. 45.5 %). According to the pathologist, this sample is distorted in the staining process, which causes the cells to have similar shapes to the cancerous ones. On the contrary, the MSHT has identified a few positive samples as normal pancreatic cells due to the limited sight-field, as shown in Fig. S4. The narrow cells present ambiguous semantic information, which confuses MSHT in classification. Besides, the complex stained background also decreases the clarity in identifying the distribution of cancerous pancreatic cells. The misclassified samples indicate the interpretability in another way. Such samples are inevitable in the clinical diagnosis, and the intaking of these samples increases the generalizability of the deep learning models.

TABLE SI
EXPERIMENTAL RESULTS OF MODELS ON THE PANCREATIC ROSE DATASET

| *Model* | *Train Acc (%)* | *Validate Acc (%)* | *Test Acc (%)* | *Train F1-score (%)* | *Validate F1-score (%)* | *Test F1-score (%)* |
|---|---|---|---|---|---|---|
| *VGG-16* | **97.70** | 93.87 | 94.92 | **96.79** | 91.43 | 92.95 |
| *VGG-19* | 96.31 | 93.81 | 94.83 | 94.82 | 91.37 | 92.78 |
| *ResNet50* | 97.58 | **94.75** | 95.02 | 96.63 | **92.76** | 93.12 |
| *Efficientnet_b3* | 93.54 | 92.04 | 93.29 | 90.88 | 88.69 | 90.51 |
| *Efficientnet_b4* | 86.52 | 88.36 | 90.91 | 80.31 | 82.86 | 86.94 |
| *Inception V3* | 94.70 | 93.69 | 93.84 | 92.58 | 91.24 | 91.49 |
| *Xception* | 95.17 | 93.58 | 94.69 | 93.20 | 90.95 | 92.55 |
| *Mobilenet V3* | 91.92 | 92.10 | 93.44 | 88.60 | 88.76 | 90.80 |
| *ViT* | 97.06 | 94.28 | 94.50 | 95.87 | 92.03 | 92.37 |
| *DeiT* | 94.53 | 93.25 | 94.52 | 92.30 | 90.57 | 92.42 |
| *Swin Transformer* | 95.74 | 94.70 | 94.92 | 94.02 | 92.54 | 93.03 |
| ***MSHT(Ours)*** | 97.53(-0.17 %) | 94.37(-0.38 %) | **95.68 (+0.66 %)** | 96.55(-0.24 %) | 91.98(-0.78%) | **93.94 (+0.82 %)** |